# Notes on derivation of streamline fields using artificial neural networks for automatic simulation of material forming processes

H. Goodarzi Hosseinabadi [a,*]

[a] Department of Materials Science and Engineering, Sharif University of Technology, Azadi Ave., P.O.Box 11155-9466, Tehran, Iran

## Abstract

Here I introduce an automatic approach to determine the material flow patterns during deformation process using artificial neural networks (ANN). Since deriving and calibrating complex mathematical models for prediction of power requirements in each individual deformation process is inconvenient, the generality of using streamline field method has been limited. I propose an automatic approach to build and calibrate streamlines with ANN. The coordinates of specific points within the deformation region were used as input and the stream function values on the points were used as output dataset in ANN training algorithm. A specific neural network architecture was then implemented to predict the flow patterns of the deforming body by an ANN-based streamline equation. At the next step, the upper bound theorem was incorporated to estimate the force and power requirements for equal channel angular extrusion (ECAE) of a composite system under hot deformation. For verification of the results, a finite element software was utilized in parallel to investigate the accuracy of the proposed approach for estimation of force requirements. The predicted force requirements under various temperature ranges with ANN based technique were consistent with that of the finite element predictions which demonstrates the accuracy of the proposed approach. The proposed approach may be appropriate for fast simulation of a wide range of steady state material forming processes.

***Keywords:*** *Upper bound theorem; Artificial neural networks; Finite elements; Streamline field; Material forming processes.*

[*] ArXiv preprint

## 1. Introduction

One of the severe plastic deformation (SPD) techniques, which is currently considered as the most promising tone for industrializing, is equal channel angular extrusion (ECAE). This process is renowned for its simplicity, homogeneity after deformation, capabilities of commercialization and applying high strain values to the ECAEed material. Complexity of the material flow pattern during ECAE can be accounted as another characteristic of this process. In recent years, the finite element method (FEM) has been widely utilized to simulate the flow patterns of the material during deformation [1-5]. However, the large amount of computational costs due to time-consuming solution of material flow equations, have motivated the industry to save time and money by using alternative approaches, such as the upper bound techniques [6]. Among those, the streamline methods have shown acceptable contribution to this aim [7, 8]. By the way, deriving and adjusting an appropriate analytical model to predict the flow patterns for each individual process has been a challenge during past years [9-13]. The inconvenience of calculating complex mathematical equations has been known as one of the limitations of the upper bound methods, since each proposed model works on specific boundary condition and for a special process. Thus, the generality of the solution and method is largely limited [11].

The present study is an attempt to introduce a novel, fast and suitable method to calculate the deformation field with the aid of artificial neural networks (ANN) at a reasonable expense of lower precision. The proposed method is useful for a wide range of steady state processes without spending long time to derive appropriate mathematical models for each individual process and to adjust the geometry of the deformation region. On the other hand, it is likely that the proposed idea after maturation can put an end to one of the major limitations of simulation works that implement the upper bound theorem. With this aim, the concept of ANN-based streamline field is introduced here for the first time and it is applied to predict the complex flow pattern occurring in a complicated metal forming process, i.e. ECAE under hot temperature conditions. The accuracy of modeling results are evaluated by running a finite element code for the exact deformation process. The agreement obtained between the results validates the applicability of the proposed method. The smaller computational cost of the present method compared to the FE method highlights the novelty and applicability of the introduced method.

## 2. Deformation model

### *2. 1. Stream function configuration*

It is shown that for numerous metal forming processes, the deformation field can be modeled by the streamline method [7, 8]. In this regard, a scalar function called "the stream function" is defined and accordingly, the velocity field can be distinguished by the stream function derivatives. For instance, in case of two-dimensional Cartesian coordinates, the stream function describes the velocity field as follows:

$$V_X = \frac{\partial \Psi}{\partial Y} \tag{1}$$

$$V_Y = -\frac{\partial \Psi}{\partial X} \tag{2}$$

where $\Psi(X,Y)$ demonstrates the stream function and $V_X$ and $V_y$ denote the velocity components along the X and Y directions, respectively. An interesting utility of the above definition is to maintain the volume constancy principle during the plastic deformation. For steady state processes such as ECAE, where the stream lines coincide with flow path lines, the steady state velocity distribution can be obtained by stream function values which:

1. Satisfy the boundary conditions of the stream function,

2. Take identical values on each streamline,

3. Be compatible with nature of the material flow.

As discussed before, many investigators have made their attempts to propose analytically suitable functions as the stream function for various metal forming operations. It should be noted that in this manner not only a large number of mathematical models are published [9-13], but also a considerable amount of complexity and calculus is needed by an investigator before he or she can interpret the deformation field expeditiously. The present work makes an attempt to introduce ANN-based stream functions that can be automatically derived by ANN and then satisfy the above-mentioned requirements utilizing a feed-forward neural network. In this regard, a neural network is trained to interpolate the stream function values at different points. To this aim, adequate number of points with known stream function values are considered on the die profile as well as within the steady state region of the die to make the neural network capable of recognizing the steady behavior. The geometry of performing ECAE on a billet of square cross section together with the distribution of the points with known stream function values are illustrated in Figure 1. The method of calculating stream function values at these points will be described in later section. Since the mentioned process takes place under a plane strain condition, the points and relative calculations are considered on two-dimensional plane. In Figure 1, $R_{if}$ indicates the fillet radius of the inner die wall and $R_{of}$ denotes the outer fillet radius dependent on the formation of dead metal zone (DMZ). In the present research, the dimensionless parameter $R_{of} / l$ implies the DMZ configuration where '$l$' denotes the billet width. It is worth noting that, supposing analogous frictional conditions between the billet and all walls of the die channel, the DMZ profile is as shown in Figure 1. As mentioned earlier, in order to make the network capable of interpolating the steady state flow, some of points inside the steady state region were also needed to be considered in training dataset. Figure 1 also illustrates the volume of steady region and the points that utilized in this regard. The value of stream function for all of the points inside the "ABCD" and "GHIJ" regions are obtained by the following equations:

$$V_X = \frac{\partial \Psi}{\partial Y} = 0 \, , \; V_Y = -\frac{\partial \Psi}{\partial X} = -V_0 \;\rightarrow\; \Psi_{at\,\text{"ABCD"}} = V_0 \, X \tag{3}$$

$$V_X = \frac{\partial \Psi}{\partial Y} = V_0 \, , \; V_Y = -\frac{\partial \Psi}{\partial X} = 0 \;\rightarrow\; \Psi_{at\,\text{"GHIJ"}} = V_0 \, Y \tag{4}$$

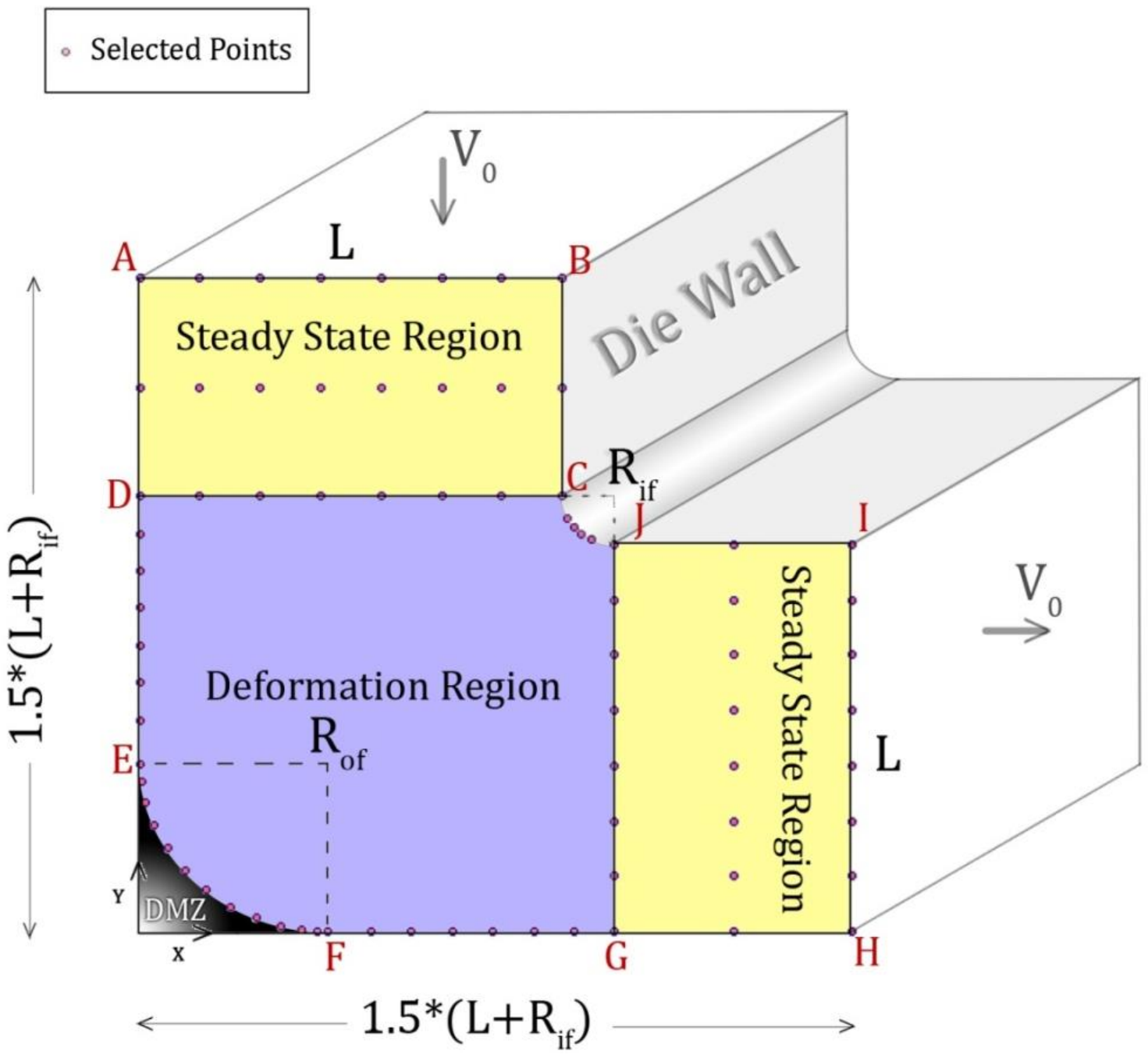

*Figure 1. Schematic geometry of ECAE process (the illustrated points that carry certain stream function values, form the neural network training dataset).*

The value of stream function for the points alongside “DE” and “FG” sections are achieved according to the following:

$$V_X = \frac{\partial\Psi}{\partial Y}=0,\ V_Y = -\frac{\partial\Psi}{\partial X} = -V_0 \ \rightarrow\ \Psi_{on\ "DE"} = V_0\ X\,|_{(X=0)} = 0 \ \rightarrow\ \Psi_{on\ "DE"} = 0 \tag{5}$$

$$V_X = \frac{\partial\Psi}{\partial Y}=V_0\ ,\ V_Y = -\frac{\partial\Psi}{\partial X} = 0 \ \rightarrow\ \Psi_{on\ "FG"} = V_0\ Y\,|_{(Y=0)} = 0 \ \rightarrow\ \Psi_{on\ "FG"} = 0 \tag{6}$$

Since “EF” segment as a DMZ boundary demonstrates a streamline, the stream function must return a constant value on this line, i.e. $\Psi_{on\ "EF"} = cte$. The exact value of this constant can be recognized using the stream function value at point “E” that reads:

$$\Psi_{on\ "EF"} = cte\ \ ,\ \Psi_{on\ "E"} = 0 \rightarrow\ \Psi_{on\ "EF"} = 0 \tag{7}$$

Similarly, since “CJ” denotes the inner fillet of the die, the stream function will take a constant value on this section, thus:

$$\Psi_{on\ "CJ"} = cte \quad , \quad \Psi_{on\ "C"} = V_0\ X\,|_{(X=L)} = V_0\ L \rightarrow \quad \Psi_{on\ "CJ"} = V_0\ L \tag{8}$$

### *2. 2. Artificial neural network (ANN) model*

After having the prescribed streamline values on the selected points, in the next step a stream function should be interpolated throughout the deformation zone that satisfies the prescribed data. In contrary with previous publications that introduce an explicit analytical expression as the stream function, the strategy here is to present the function based on ANNs. This idea emerges from the proven fact that a neural network including at least one sigmoid hidden layer, as well as including linear output layer and biases, is capable of being traind for approximating any given function with a finite number of discontinuities [14].

In the present work, the amounts of stream function at selected points with calculated stream values are used as the output and the points coordinates are used as the input training data for the neural network. Inspiring from a previously proposed network [15], a feed forward neural network [16] including two hidden layers and one output layer in its architecture is constructed using the Matlab$^{TM}$ software. The first layer consisted of 3 neurons which use the "tansig" transfer function [14] to operate on the normalized input dataset. The second layer utilizes 2 neurons, each one employs the "logsig" transfer function [14] to operate on the first layer outcomes. The last layer, constructed in compliance with the output dataset, utilized a linear transfer function. A back propagation (BP) training scheme was then applied to improve the performance of the network at each iteration. Since the "trainbr" BP training algorithm [14] was employed in the present work, the concluded network could be accounted as a Bayesian Regularized (BR) network. This type of network which combined the quick optimization algorithm of Levenberg–Marquardt and the idea of Bayesian framework [17] at the neural network would be capable of controlling the complexity of the model and prohibit the over fitting of the training dataset [18]. On the other hand, usage of a BR network could provide a good generalized neural network, so that the specified network architecture could be applicable to a wide range of datasets. In this regard, it was possible that an almost similar architecture could be used for training of different datasets, with capability of being used for deformation processes other than ECAE. However, the generality of the proposed method needs to be comprehensively examined in a course of further works.

To train the network, the input and output vectors were randomly divided into three sets. 60% of the vectors were used to train the network and 20% were used to check the validity of the network. Training process continued as long as the training procedure reduced the network error on the validation vectors [14]. As soon as an enhancement on the network error initiated, the system memorized the trained network and the training process halted. This procedure also avoided the problem of over fitting automatically, which plagues many ANN optimization and learning algorithms. Finally, the last 20% of the vectors provided an independent dataset to test the network generalization with data that the network never encountered beforehand. Afterward, the maximum percentage of error was defined by the following equation:

$$Max.Error = 100 * max \left|\frac{\Psi_{ANN} - \Psi}{\Psi}\right| \tag{9}$$

Where, $\Psi_{ANN}$ described the value of ANN-predicted stream function and the $\Psi$ vector denoted the known stream function values on each selected point. If the maximum percentage of obtained error lied beneath the convergence criterion magnitude, the memorized neural network was employed to investigate the amount of stream function throughout the deformation zone at the next steps. Otherwise, a new training course repeated until an adequately precise network was achieved. Each repetition was called an “attempt”. The outstanding characteristic of the obtained stream function by this algorithm lies on its quickness and utility that makes it convenient for computerized-simulation of many deformation processes.

## *2. 3. Formulation derivation for Al-based composites*

### *2. 3. 1. Implementation of the upper-bound theorem*

To predict the actual power requirements and dimensions of the dead metal zone, the rate of energy consumption needs to be evaluated within the deformation region under different DMZ configurations. Then, the state with minimum rate of energy dissipation is an approximation for the real rate of energy consumption and presumably the real configuration of the dead metal zone. In other words, since the estimated rate of energy, obtained by the ANN-based kinematically admissible velocity field, is always higher than or equal to the rate of work done during the process, thus –according to the upper bound theorem [6]– the velocity field which maintains the minimum rate can be a good prediction of actual energy requirements. Accordingly, the energy consumption components is calculated by the following equations:

$$\dot{W}_{Internal} = \int \bar{\sigma}\, \dot{\bar{\varepsilon}}(LdXdY) \tag{10}$$

$$\dot{W}_{DMZ} = \int \frac{\bar{\sigma}}{\sqrt{3}} |\Delta v_t| dS_{DMZ}, \ |\Delta v_t| = \sqrt{v_X{}^2 + v_Y{}^2}|_{S_{DMZ}} \tag{11}$$

$$\dot{W}_{LateralWall} = m_{Wall} \int \frac{\bar{\sigma}}{\sqrt{3}} |\Delta v_t| dS_{LateralWall} \tag{12}$$

$$\dot{W}_{Outter\ Wall} = m_{Wall} \int \frac{\bar{\sigma}}{\sqrt{3}} |\Delta v_t| dS_{OuterWall}, \tag{13}$$

$$\dot{W}_{Inner\ Wall} = m_{Wall} \int \frac{\bar{\sigma}}{\sqrt{3}} |\Delta v_t| dS_{InnerWall}, \tag{14}$$

here $\dot{W}_{internal}, \dot{W}_{DMZ}, \dot{W}_{LateralWall}, \dot{W}_{Outter\ Wall}$ and $\dot{W}_{Inner\ Wall}$ denote the rate of power dissipation due to internal deformation and frictional shearing processes on the DMZ surface, lateral walls of the die channel and on both the inner and outer surfaces of the die channel, respectively. $v_t$, implies the tangential velocity along the corresponding surface, while $m_{Wall}$ demonstrates the friction factor between the billet and the walls of die channel. Table 1 shows the temperature dependency of the friction factor that is used for hot processing of AA6061-10%SiC$_{(p)}$ composite.

*Table 1. Variations of the friction factor (m) between AA6061_10%SiC$_{(p)}$ composite and tool steel [15]*

| Temperature ($^o$C) | 300 | 400 | 450 | 500 |
|---|---|---|---|---|
| sticking friction factor (*m*) | 0.2 | 0.3 | 0.35 | 0.4 |

The illustration of surfaces such as $S_{DMZ}$, $S_{LateralWall}$, $S_{OuterWall}$ and $S_{InnerWall}$ are illustrated in Figure 2. In this research, reduced integration scheme was employed to calculate the value of power components on each element. As is obvious, performing numerical integration on each of these power components is strongly dependent on the evaluation of the material flow stress. Since in this research hot processing of AA6061_10% $SiC_{(p)}$ composite is considered, the next challenge would be establishing a relationship capable of considering the combined effect of strain rate hardening and temperature softening on the material flow stress.

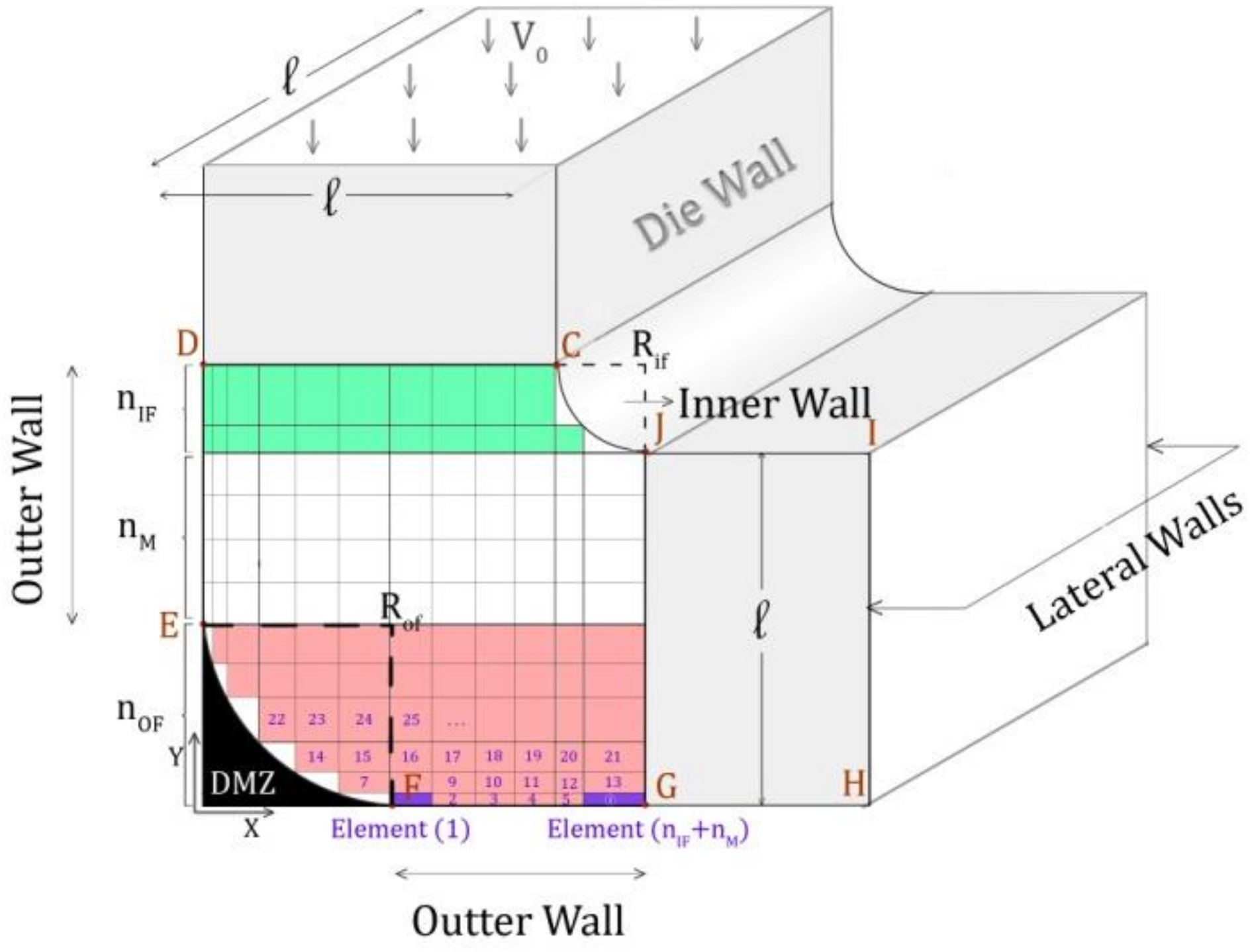

*Figure 2. Illustration of the surfaces and the mesh system adopted for ECAE process.*

It is worth noting that the effect of strain hardening on the material flow stress at the employed range of temperature and strain rate is negligible [15]. Fortunately, previous investigators [19] proposed the required constitutive equation by:

$$\bar{\sigma}[MPa] = \frac{1}{\alpha} * sinh^{-1}\left[\left(\frac{\dot{\varepsilon}}{A}\right)^{\frac{1}{n}} * exp\left(\frac{Q}{n*R*(T+273)}\right)\right] \quad (15)$$

In the above equation, $\dot{\varepsilon}$ and $T$ denote the strain rate ($s^{-1}$) and temperature ($^{o}C$), while the other material dependent parameters carry constant magnitudes as summarized in Table 2.

*Table 2. Coefficients of the constitutive equation to define hot behavior of AA6061_10%$SiC_{(p)}$ [15]*

| n | A | Q (kJ/mol) | $\alpha$ ($MPa^{-1}$) | R(J/mol $^{o}C$) |
|---|---|---|---|---|
| 3.165 | 3.005e17 | 287.7 | 0.055 | 8.314 |

Since current paper considers isothermal deformation condition for the billet, where the slow speed of the process lets sufficient heat exchange between the billet and the die, the temperature distribution inside the billet can be supposed constant during the process. The accuracy of the above assumption is investigated individually using a finite element simulation.

Since the amount of next variable, i.e. strain rate, strongly depends on the local velocity gradients, firstly the velocity field should be recognized to obtain the strain rate magnitude. Replacing the partial derivatives by the corresponding central differences [20], the velocity field is obtained in terms of the proposed stream function as:

$$V_X = \frac{\partial \Psi(X,Y)}{\partial Y} = \frac{\Psi(X,Y+\Delta Y)-\Psi(X,Y-\Delta Y)}{2\Delta Y} \quad (16)$$

$$V_Y = -\frac{\partial \Psi(X,Y)}{\partial X} = -\frac{\Psi(X+\Delta X,Y)-\Psi(X-\Delta X,Y)}{2\Delta X} \quad (17)$$

Here, $\Delta X$ and $\Delta Y$ denote the average dimensions of the corresponding element. Accordingly, the strain rate components are attainable using the following expressions:

$$\dot{\varepsilon}_X = \frac{\partial V_x}{\partial X} = \frac{\partial^2 \Psi}{\partial X \partial Y} = \frac{\Psi(X+\Delta X,Y+\Delta Y)+\Psi(X-\Delta X,Y-\Delta Y)-\Psi(X-\Delta X,Y+\Delta Y)-\Psi(X+\Delta X,Y-\Delta Y)}{4\Delta X \Delta Y} \quad (18)$$

$$\dot{\varepsilon}_Y = \frac{\partial V_Y}{\partial Y} = -\frac{\partial^2 \Psi}{\partial Y \partial X} = -\dot{\varepsilon}_X \quad (19)$$

$$\dot{\varepsilon}_{XY} = \frac{1}{2}\left(\frac{\partial V_X}{\partial Y} + \frac{\partial V_Y}{\partial X}\right) = \frac{1}{2}\left[\frac{\partial^2 \Psi}{\partial Y^2} - \frac{\partial^2 \Psi}{\partial X^2}\right] = \frac{1}{2}\left[\frac{\Psi(X,Y+\Delta Y)+\Psi(X,Y-\Delta Y)-2\Psi(X,Y)}{\Delta Y^2} - \frac{\Psi(X+\Delta X,Y)+\Psi(X-\Delta X,Y)-2\Psi(X,Y)}{\Delta X^2}\right] \quad (20)$$

Eventually, equivalent strain rate on each desired coordinate can be approximated as:

$$\dot{\bar{\varepsilon}} = \sqrt{\frac{2}{3}\left(\dot{\varepsilon}_X^2 + \dot{\varepsilon}_Y^2 + 2\dot{\varepsilon}_{XY}^2\right)} \quad (21)$$

Now, with substitution of the elemental strain rate and temperatures in equation (15), the flow stress can be determined on each local element. Accordingly, the numerical integration was calculated and the energy dissipation components obtained in the deformation region.

**2. 3.2 Solution procedure for Al-based composites**

The deformation zone was meshed with quadrilateral iso-parametric elements, while finer elements were evenly distributed near the die wall and fillets, as shown in Figure 2. The number of elements were deliberated such that the analysis response does not oscillate significantly with higher mesh refinement. Finally, the upper bound theorem was used to investigate the mechanical behavior of the material during ECAE process.

Based upon the explanations provided above, firstly a qualified neural network was trained to interpolate the values of the stream function on approximately 90 points situated on the boundary surfaces and also within the steady state region, as discussed earlier. The training attempts were repeated until an appropriate trained network with maximum error of less than convergence criterion obtained. After each successful training, a visual illustration of the streamline contour was provided automatically to display any possibilities of undesired behaviors of neural network and to ensure the accuracy of the network adjacent to the boundaries by visual checking in the current version of the work as a preliminary version. The visual checking accelerates finding of those flow patterns that are closer to the actual flow pattern. However, the visual checking can be eliminated from the above procedure when, a convergence criterion be considered with lower values and instead, more repetitive runs be followed to find an appropriately precise ANN-based stream function. According to the hill's minimum principle, the actual flow pattern results in the minimum power consumption [6], and thus, the network -among the repetitive runs- that results in minimum amount of power consumption is a candidate to illustrate the nearest pattern to the actual flow pattern. Indeed, here the additional computations with inaccurate networks can be avoided by visual checking of the flow patterns and can effectively eliminate the large computational costs required for seeking between networks.

Lastly, the velocity field and strain rate field were predicted at the deformation region using the achieved ANN-based stream function. Assuming isothermal conditions, the flow stress distribution was calculated at each local element. Note that the rate of energy consumption for various DMZ configurations was computed employing the reduced integration scheme upon integrations introduced by Equations (10) to (14), and according to the upper bound theorem, the configuration with minimum rate of energy dissipation can be the closest delineation to the actual DMZ configuration.

## *2. 4. Formulation derivation for Mg-based composites*

### *2. 4. 1. Implementation of the upper-bound theorem*

In case of AZ31 alloy, which shows dynamic recrystallization during the deformation and the derivation of formulations needs extra effort, one may use a constant of 0.3 as the sticking friction factor ($m_{Wall}$) between 200 to 300°C for simplicity.

Then, similar equations showed in Eq. 10-14 solved using the center-point integration scheme to calculate the amount of presented energy components on each element. Calculation of all energy components is strongly dependent to the evaluation of material flow stress. Since in this study, hot processing of AZ31 magnesium alloy is considered, the next challenge would be demonstrating a relationship capable of considering the combined effect of hardening and softening during dynamic recrystallization (DRX) phenomenon during AZ31 deformation. Fortunately, previous investigators [22] proposed the required constitutive relationship that takes into account the accumulative influence of temperature, strain rate and strain by equation 22:

$$\sigma = exp\left[\psi(\varepsilon - \varepsilon_p)^2 \ln \xi\varepsilon + \ln \sigma_p\right] \quad (22)$$

where $\varepsilon_p$ and $\sigma_p$ demonstrate the peak strain and the peak stress of the material when DRX takes place during the deformation. In previous equation, $\xi$ is a constant parameter while $\psi$ depends on the deformation temperature and strain rate via the 'Z' parameter in the form:

$$\psi = exp\left(\frac{\ln Z - B_1}{B_2}\right) \quad (23)$$

The peak strain and stress variables can be written as:

$$\sigma_p = \frac{1}{\alpha} \ln\left\{\left(\frac{Z}{A}\right)^{\frac{1}{n}} + \left[\left(\frac{Z}{A}\right)^{\frac{1}{n}} + 1\right]^{\frac{1}{2}}\right\} \tag{24}$$

$$\varepsilon_p = exp\left(\frac{\ln Z - C_1}{C_2}\right) \tag{25}$$

In the above relationships, Z implies to the Zener-Hollomon parameter and obtains by the next equation:

$$Z = \dot{\varepsilon}\, exp\left(\frac{Q}{R\,T}\right) \tag{26}$$

While T is the absolute temperature in Kelvin and R is the universal gas constant (8.314 J/mol/ K). The other parameters in the expressed equations are 9 independent constants that are obtained by the least-square method for the AZ31 alloy [22] in the range of 523 to 673$^o$K temperatures, 0.001 to 1 $s^{-1}$ strain rates and experiencing less than $1_{mm/mm}$ strains. Table 1, summarizes the magnitude of mentioned coefficients.

Table 1. Coefficients of the constitutive equations to define hot deformation of AZ31 magnesium alloy **[22]**.

| n | A | Q (J/mol) | A | ξ | $B_1$ | $B_2$ | $C_1$ | $C_2$ |
|---|---|---|---|---|---|---|---|---|
| 0.15 | 1.18 | 141780 | 2428153.38 | 84.33 | 13.37 | -7.32 | 25.30 | 11.35 |

As can be inferred from the above formulations, the flow stress of the material is just dependent to the amount of three variables: temperature, strain rate and strain.

Since the current study considers isothermal deformation of strip samples, where a substantial contact surface exist between the strip and the die, the temperature of the strip during the ECAE process can constantly be assumed equal to the die temperature.

The amount of the second variable, i.e. strain rate, is strongly dependent to the local coordinates. In order to obtain the strain rate filed, the velocity field should firstly be recognized. Replacing the partial derivatives by the corresponding central differences [23], the velocity field furnishes in terms of the proposed stream function as:

$$V_X = \frac{\partial \Psi(X,Y)}{\partial Y} = \frac{\Psi(X,Y+\Delta Y)-\Psi(X,Y-\Delta Y)}{2\Delta Y} \tag{27}$$

$$V_Y = -\frac{\partial \Psi(X,Y)}{\partial X} = -\frac{\Psi(X+\Delta X,Y)-\Psi(X-\Delta X,Y)}{2\Delta X} \tag{28}$$

Here, $\Delta$X and $\Delta$Y denote the average dimensions of the corresponding element. Accordingly, the strain rate components are attainable using the next expressions:

$$\dot{\varepsilon}_X = \frac{\partial V_x}{\partial X} = \frac{\partial^2 \Psi}{\partial X \partial Y} = \frac{\Psi(X+\Delta X,Y+\Delta Y)+\Psi(X-\Delta X,Y-\Delta Y)-\Psi(X-\Delta X,Y+\Delta Y)-\Psi(X+\Delta X,Y-\Delta Y)}{4\Delta X\Delta Y} \tag{29}$$

$$\dot{\varepsilon}_Y = \frac{\partial V_Y}{\partial Y} = -\frac{\partial^2 \Psi}{\partial Y \partial X} = -\dot{\varepsilon}_X \tag{30}$$

$$\dot{\varepsilon}_{XY} = \frac{1}{2}\left(\frac{\partial V_X}{\partial Y} + \frac{\partial V_Y}{\partial X}\right) = \frac{1}{2}\left[\frac{\partial^2 \Psi}{\partial Y^2} - \frac{\partial^2 \Psi}{\partial X^2}\right] = \frac{1}{2}\left[\frac{\Psi(X,Y+\Delta Y)+\Psi(X,Y-\Delta Y)-2\Psi(X,Y)}{\Delta Y^2} - \frac{\Psi(X+\Delta X,Y)+\Psi(X-\Delta X,Y)-2\Psi(X,Y)}{\Delta X^2}\right] \tag{31}$$

At last, the equivalent strain rate on each desired coordinate can be approximated by the next equation:

$$\dot{\bar{\varepsilon}} = \sqrt{\frac{2}{3}\left(\dot{\varepsilon}_X^2 + \dot{\varepsilon}_Y^2 + 2\dot{\varepsilon}_{XY}^2\right)} \tag{32}$$

As described earlier, in order to perform numerical integration on the power dissipation equations, the flow stress should be evaluated. In this regard, dependency of the flow stress on two of the deformation variables, i.e. temperature and strain rate, described. By the way, the next challenge that would be discussed is demonstration of the third variable: the strain that is induced on each individual deformed element.

Since the strain that each particle experiences is absolutely dependent to the path line that the particle travels during the deformation, the strain history of each node must be calculated using the related streamline. In this regard, the streamline that passes from each nodal coordinate calculated using the presented velocity vectors and divided into adequately small segments. Afterward, the strain rate and velocity vectors calculated on the center of each segment using the respective stream function. Then the below equation used to carry out a numerical integration to estimate the strain history on the nodal coordinate:

$$\bar{\varepsilon} = \int \dot{\bar{\varepsilon}}\, dt = \int \dot{\bar{\varepsilon}}\, \frac{dv}{dl} \tag{33}$$

In this equation, 'l' denotes the segment length and 'v' expresses the average speed of particle within the segment. Consequently, the flow stress and thus the power dissipation components are accessible through the deformation region.

#### *2. 4.2 Solution procedure for Mg-based composites*

The deformation zone meshed with quadrilateral iso-parametric elements, and finer elements evenly distributed near the die wall and fillets. The number of elements deliberated such that the analysis response does not oscillate significantly with more mesh refining. Finally, the upper bound theorem used to investigate the mechanical flow behavior of the material during performing ECAE process on AZ31 strip samples.

First, a qualified neural network trained to interpolate the transpicuous amounts of the stream function on the boundary surfaces and within the steady state region. The training attempts repeated while an accurate trained network obtained.

Using the trained ANN-based stream function, the velocity field, strain field and the strain rate field predicted at the deformation region. Assuming isothermal condition, the flow stress distribution calculated for each local element. At the next step, employing a numerical integration scheme the rate of power consumption for diverse DMZ configurations computed.

According to the upper bound theorem, the configuration with minimum rate of energy dissipation can be the closest delineation of the actual power requirements.

## 3. Finite element simulations

For verification of the force requirement estimations obtained by the proposed model, the ABAQUS/Standard$^{TM}$ software was employed. For brevity, only the formulations derived for AA6061 Al-based composite are simulated by FEM and discussed in following sections. A transient thermo-mechanical model was designed to investigate the temperature variations alongside the mechanical response of AA6061_10%SiC$_{(p)}$ composite during ECAE process. A schematic view of the FE model is shown in Figure 3.

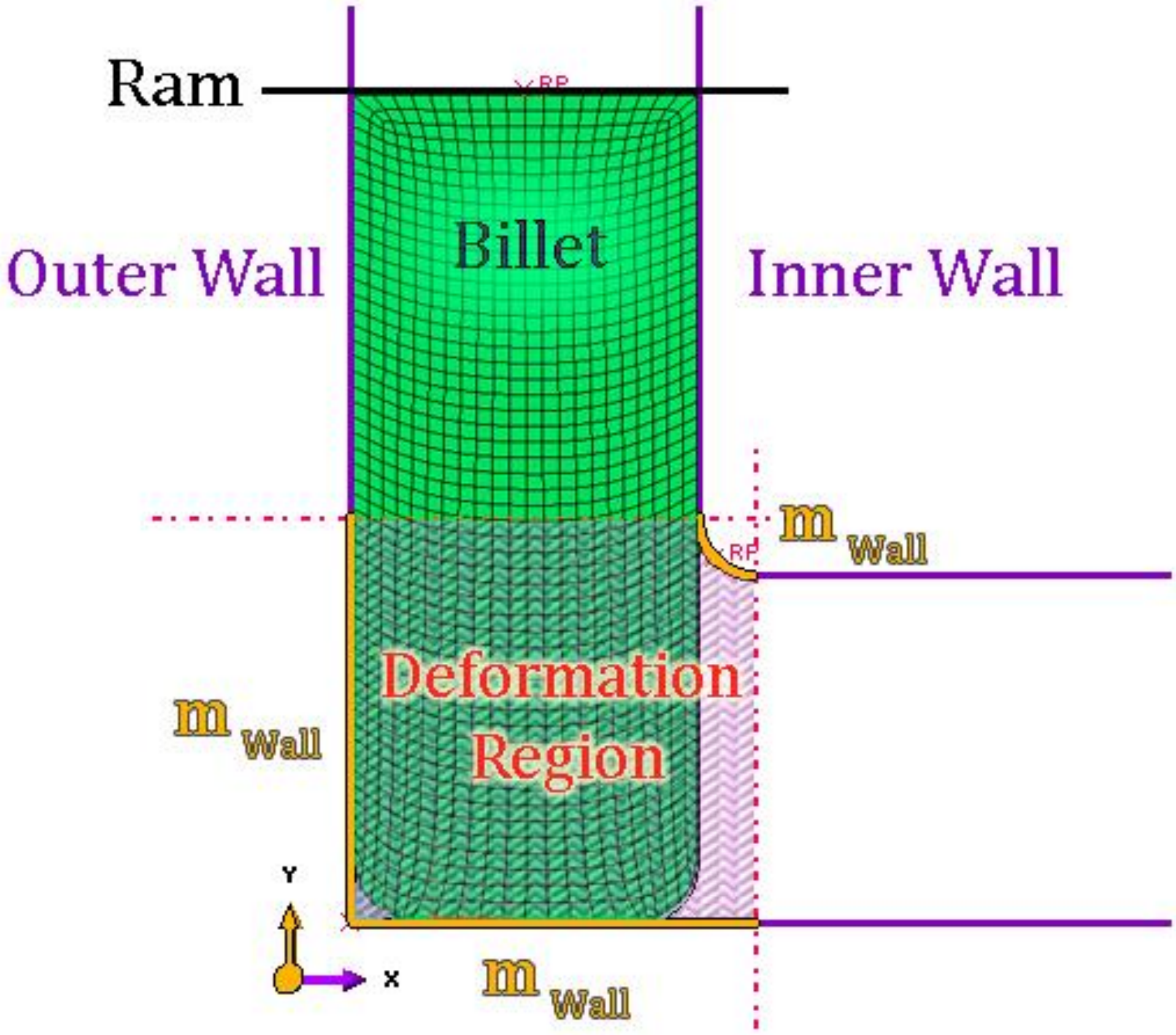

*Figure 3. Schematic view of the employed finite element model.*

Frictional conditions were selected in correspondence with the proposed streamline model to make the comparison between two models reasonable. In this regard, the friction on the walls was considered just at the hatched deformation region. A 30mm×30mm×80mm billet, was modeled with 4-node plane strain linear elements with both thermal and displacement degrees of freedom. The reduced integration algorithm together with hourglass control were implemented to simulate the temperature rise and force requirements during performing ECAE at 0.5 mm/s ram speed under 320, 400 and 450$^{o}$C temperatures. The die walls were considered isothermally rigid surfaces. Accordingly, the heat exchange between the billet and the walls was ignored to estimate the maximum possible temperature rise during the process.

**4. Results and Discussions**

Figure 4 shows a typical network performance diagram during the training procedure. Smaller "Performance" magnitude indicates smaller deviation of predictions from the training dataset, and thus means better training proficiency.

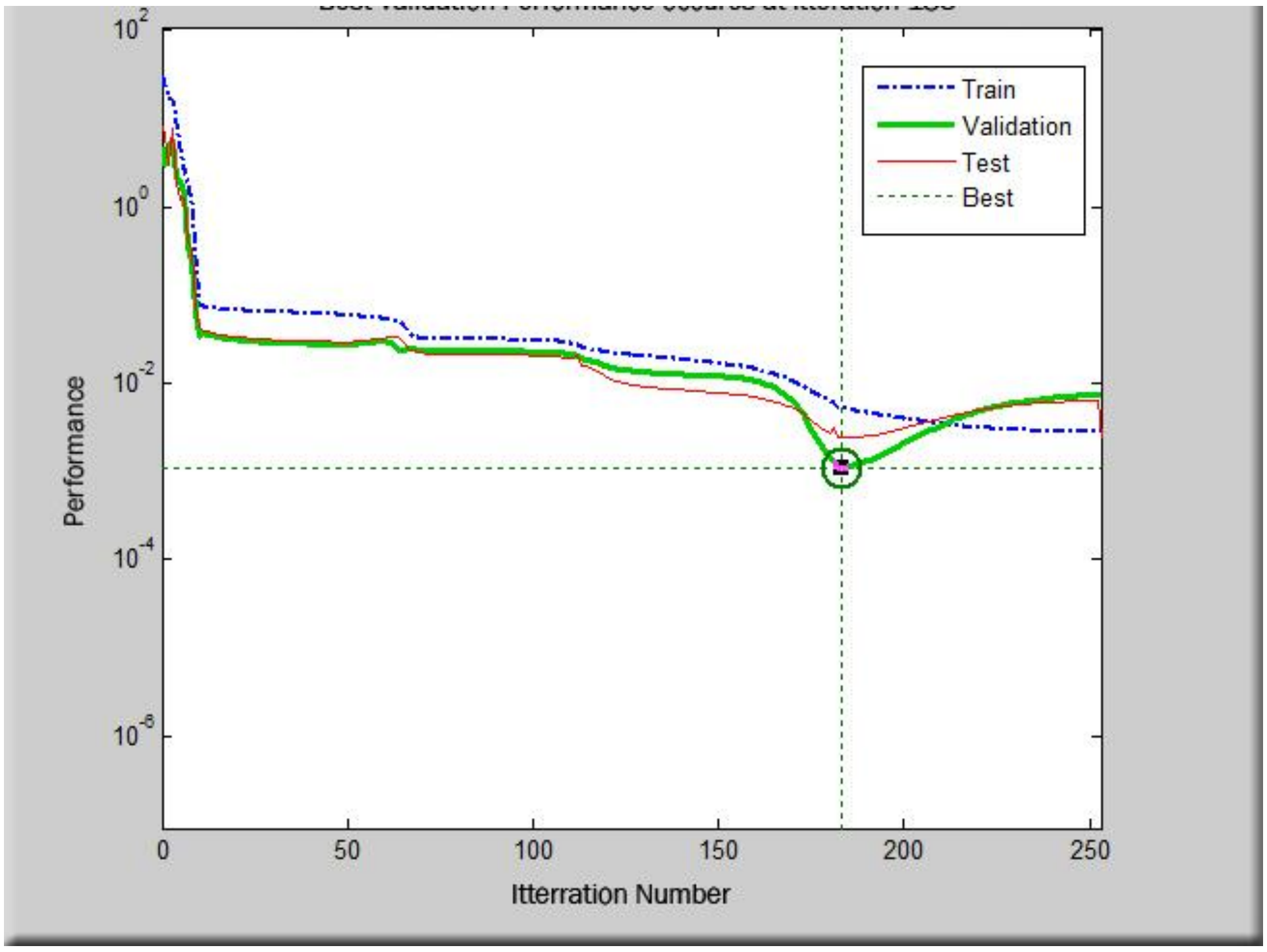

*Figure 4. Network performance variations for the three sets of data: 1) training data, 2) validating data, 3) testing data.*

As can be seen in this case, the best validation performance occurs at iteration 183 where the network initiates over fitting and the error of validation and test data increase continuously. As described earlier, the network that stores at this iteration and uses the Bayesian Regulation as training algorithm can result in an appropriate neural network for next steps.

Figure 5 illustrates the location of training points together with the quality of streamlines predicted by the proposed neural network under various DMZ configurations.

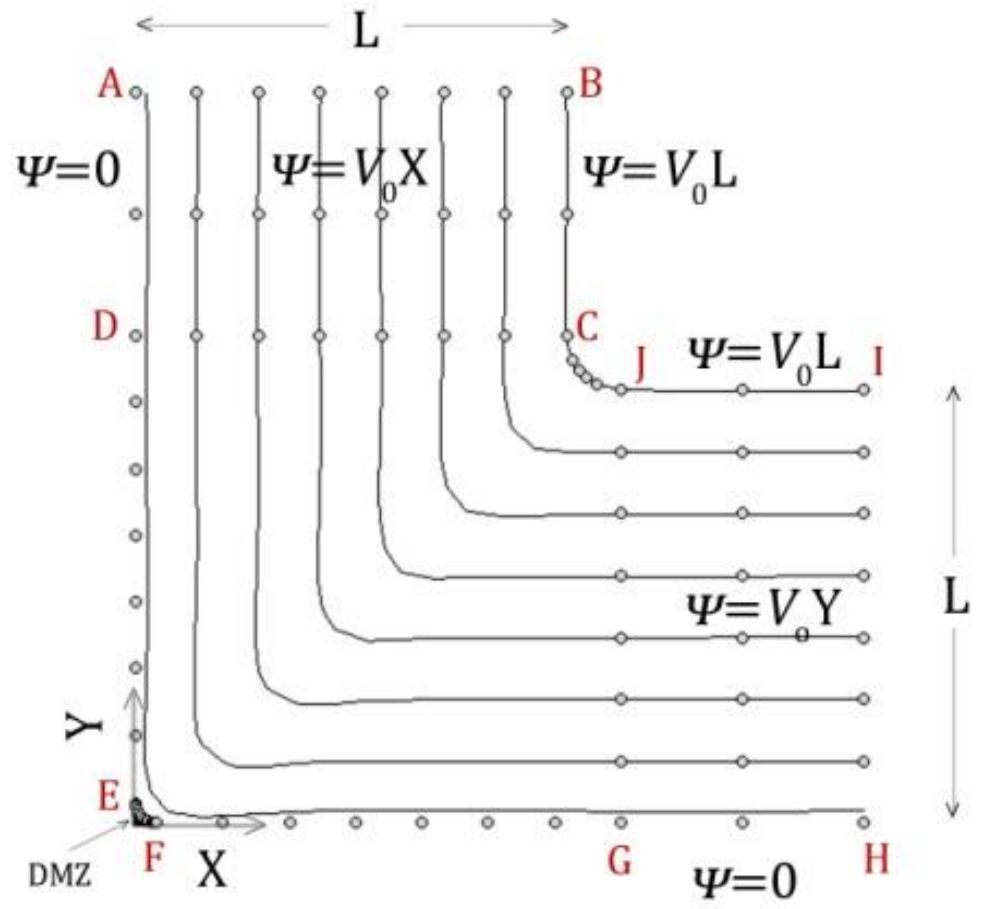

**(a)**

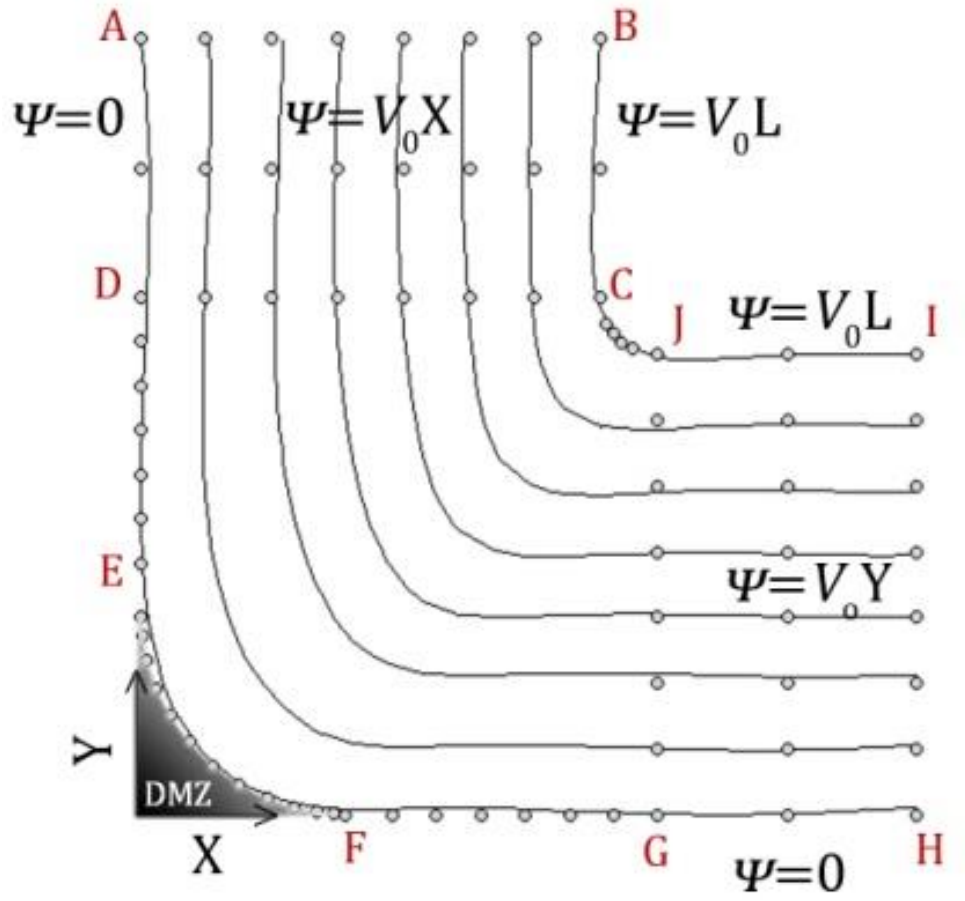

**(b)**

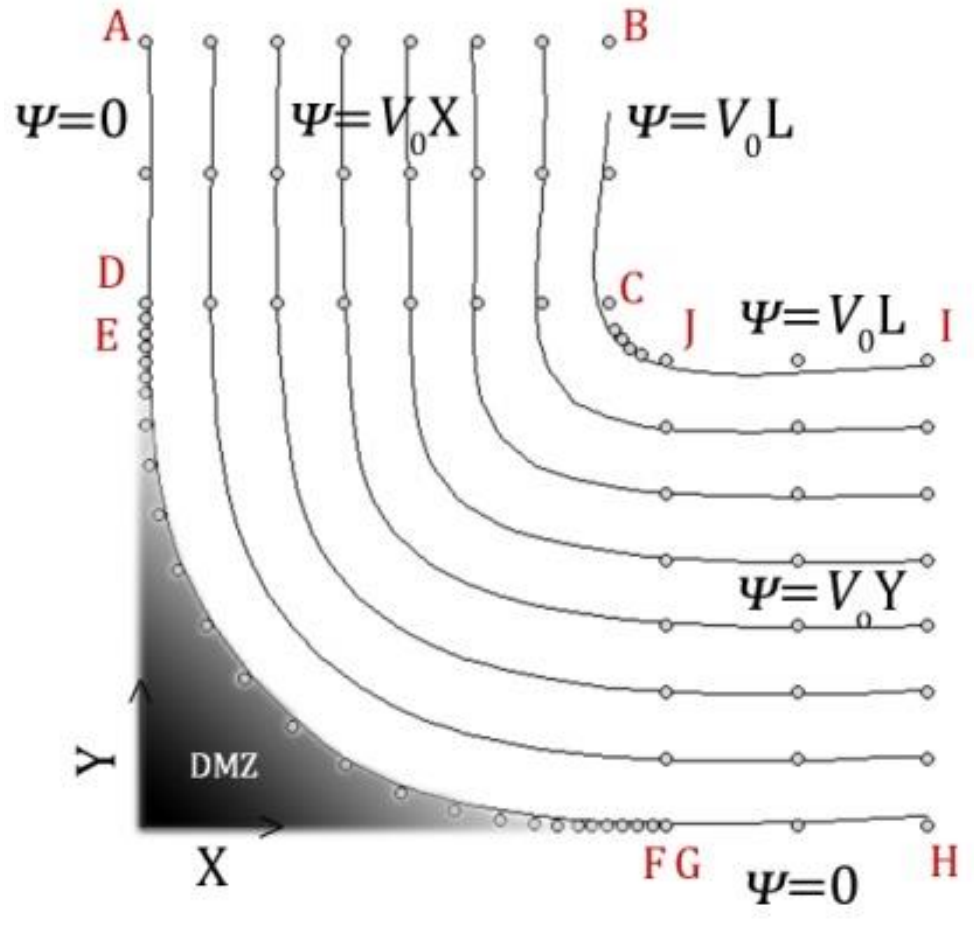

**(c)**

*Figure 5. Illustrating the quality of streamlines predicted by the trained neural network during ECAE under a)* $\frac{R_{of}}{L} = 0.05$, *(maximum error decreased to 1.71% after 5 attempts within less than 27 seconds), b)* $\frac{R_{of}}{L} = 0.45$, *(maximum error decreased to 1.23% after 2 attempts within less than 11 seconds), c)* $\frac{R_{of}}{L} = 0.9$, *(maximum error decreased to 1.1% after 2 attempts within less than 8 seconds).*

The significance of the proposed method lay on its sufficient quickness alongside its adequately precise algorithm to obtain a converged solution. In this example, the converged solution with maximum error of less than 2% was obtained within less than 30 seconds for all of the cases. Note that selecting smaller convergence criterion could result in better accuracy for the calculated streamlines, but it would be associated with slower convergence rate. In present work, the convergence criterion was chosen to be as large as 2% and simultaneously the converged streamline contours were checked visually to insure the network accuracy. In scarce situations, where the achieved contours had unacceptable quality, the network was discarded and the training procedure repeated to propose an acceptable network. However, the convergence rate for a typical feed forward neural network can strongly be dependent on the initial weight and bias matrices, a good consistency observed on the convergence rate of the simulation under various situations that occur because of the specific network architecture.

Table 3 is presented to show the consistency of the proposed model, thus the convergence quality of training algorithm is summarized for a variety of DMZ configurations. Accordingly, the number of attempts together with the history of the network's maximum error is shown in the table, as well. The table consists of two datasets concluded from two subsequent runs of the program. The difference between results of the subsequent runs emerges from the stochastic nature of the neural network. On the other hand, since the neural network considers randomly weight and bias magnitudes, the training process can experience different paths at each individual attempt and thus, an exactly similar convergence rate for different runs could not be anticipated. As can be inferred, for very small $\frac{R_{of}}{L}$ values, i.e. very small dead metal zones, the neural network needs longer times to accommodate the stream lines with the sharp flow patterns; while no meaningful difference is obtainable for larger dead metal regions (higher $\frac{R_{of}}{L}$ values). It is likely that employment of sigmoid transfer functions, i.e. 'tansig' and 'logsig' functions, at the network architecture have made the network more compatible with curved flow patterns in contrast with the sharp patterns. Anyway, the order of required time for training the network is obviously trivial in comparison with some other methods, for instance finite element approach.

*Table3. Convergence of the training procedure for proposed network at two subsequent runs*

| **First Run** | | | | | |
|---|---|---|---|---|---|
| $R_{of}/l$ (DMZ parameter) | 0.05 | 0.3 | 0.5 | 0.78 | 0.9 |
| Number of Attempts | 5 | 3 | 3 | 2 | 2 |
| Maximum Error at each attempt (%) | 3.03→ 3.03→ 3.03→ 3.03→ 1.81 | 4.04→ 2.35→ 0.64 | 5.47→ 5.47 → 1.35 | 3.27→ 1.45 | 64.3→ 1.98 |
| Elapsed time (sec) | 26.05 | 17.06 | 17.41 | 11.72 | 7.62 |
| **Second Run** | | | | | |
| $R_{of}/l$ (DMZ parameter) | 0.05 | 0.3 | 0.5 | 0.78 | 0.9 |
| Number of Attempts | 7 | 1 | 4 | 1 | 3 |
| Maximum Error at each attempt (%) | 9.33→ 2.29→ 2.29→ 2.29→ 2.29→ 2.29→ 1.62 | 0.45 | 2.74→ 2.05→ 2.05→ 1.41 | 1.62 | 3.99→ 3.86→ 1.27 |
| Elapsed Time (sec) | 47.05 | 11.67 | 21.6 | 11.57 | 8.42 |

Figure 6 compares the power consumption components obtained at different DMZ configurations.

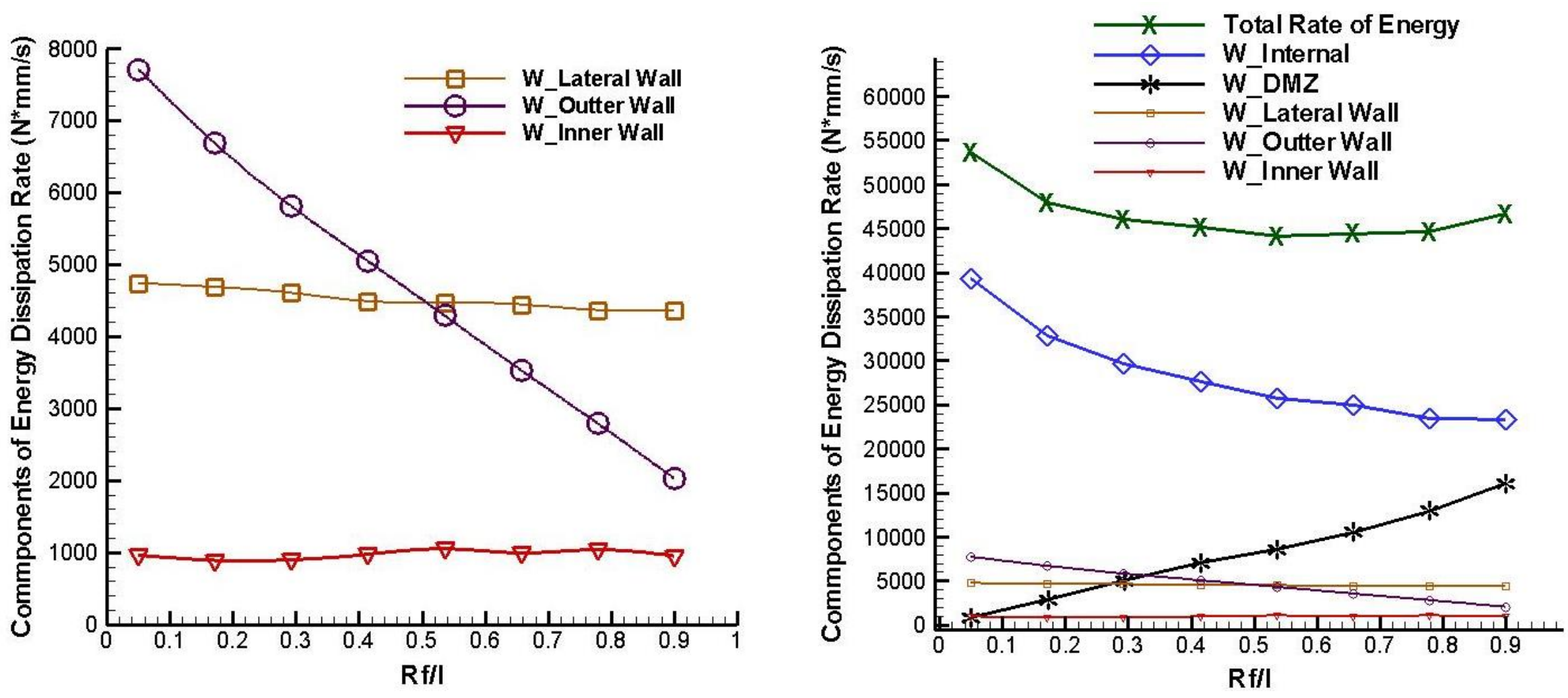

(a) (b)

*Figure 6. Comparison between rates of energy dissipation a) for die walls, b) for all of the components.*

Figure 6 (a) illustrates the frictional power dissipations near die walls. By increasing the dead metal region, the interface between the deformation region and outer wall of the die will diminish, while the inner wall interface with deformation region remain invariant. Consequently, the power dissipation decreases on outer wall and remain constant on inner wall. The lateral wall power consumption remains somewhat constant, as a consequence of spreading the region with considerable flow stress on lateral surface at the same time that the lateral interface decreases slightly by increasing dead metal region.

Figure 6 (b) compares the effect of all of the power components. As is obvious, the internal power and the DMZ interface power consumptions form the most effective components and the effect of other frictional components is negligible. The competition between increase of power dissipation due to enhancement of DMZ interface and decreasing of internal power dissipation leads to an optimum value for the total rate of energy dissipation. Consequently, the force and pressure requirements can be calculated for each condition. According to the upper bound theorem, the optimum that emerges among numerous kinematically admissible velocity fields can demonstrate the upper bound of the actual force requirements. In the present work, a finite element analysis was employed to confirm the predicted force requirements. Figure 7 is provided in this regard to compare the force predictions of the proposed model and finite element simulation. As is shown, the ANN-based streamline prediction can be taken into account as a valuable upper bound approach for the force requirements under various temperatures.

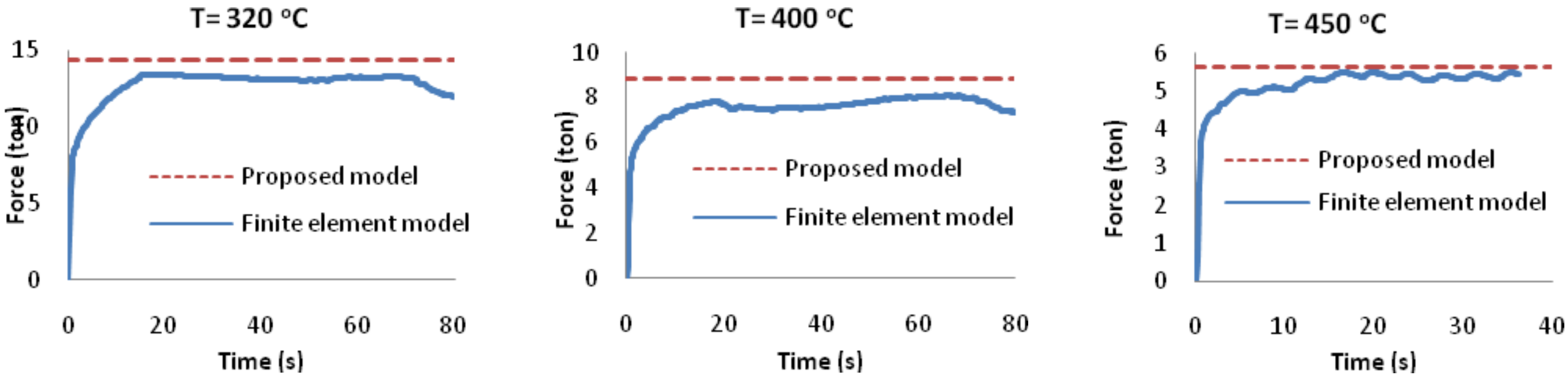

(a) (b) (c)

*Figure 7. Verification of the force predicted by the proposed model and by FEM during ECAE at ram speed of 0.5mm/s and under temperature of a) 320$^o$C, b) 400$^o$C, c) 450$^o$C.*

Note that the temperature distribution that was investigated by the finite element showed that the maximum temperature rise due to heat of deformation during ECAE at 0.5mm/s ram speed and initial temperatures of 320$^o$C and 400$^o$C are slight and equal to 15 and 7$^o$C, respectively. Since in this work, the heat exchange between the billet and isothermal die walls ignored, the actual temperature rise can be less than the aforementioned amounts. By the way, the early model assumption of isothermal condition that was employed to calculate the material flow stress during the process is reasonable.

Figure 8 shows the pressure requirements that raises from the discussed power consumptions at different temperatures.

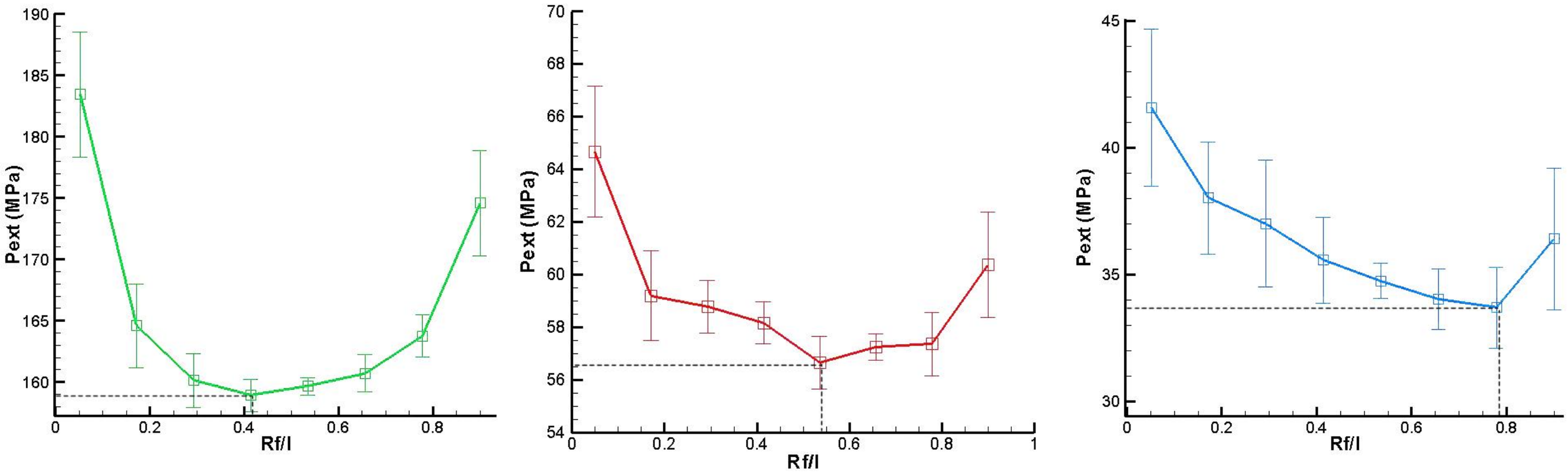

(a) (b) (c)

*Figure 8. Illustration of the required external pressure variations at temperature of a) 320$^o$C, b) 400$^o$C, c) 500$^o$C.*

As can be observed, the pressure requirements decrease extensively with rising the process temperature. The indicated minima show the approximation of upper bound theorem for actual pressure requirements. With ignoring the concerns related to the error bar variations (that is inevitable in this approach), it seems that by increasing the process temperature, the optimum pressure occurs at higher “$R_{of}/l$” parameters. Therefore, larger dead metal regions are predicted to be formed at higher temperatures. Since the friction factor increases by the temperature, more sticky condition is anticipated at higher temperatures. Hence, the prediction of growth in dead metal region seems to be sensible.

Figure 9 illustrates the predicted contours for very small and optimized DMZ configurations.

Small DMZ

Optimized DMZ

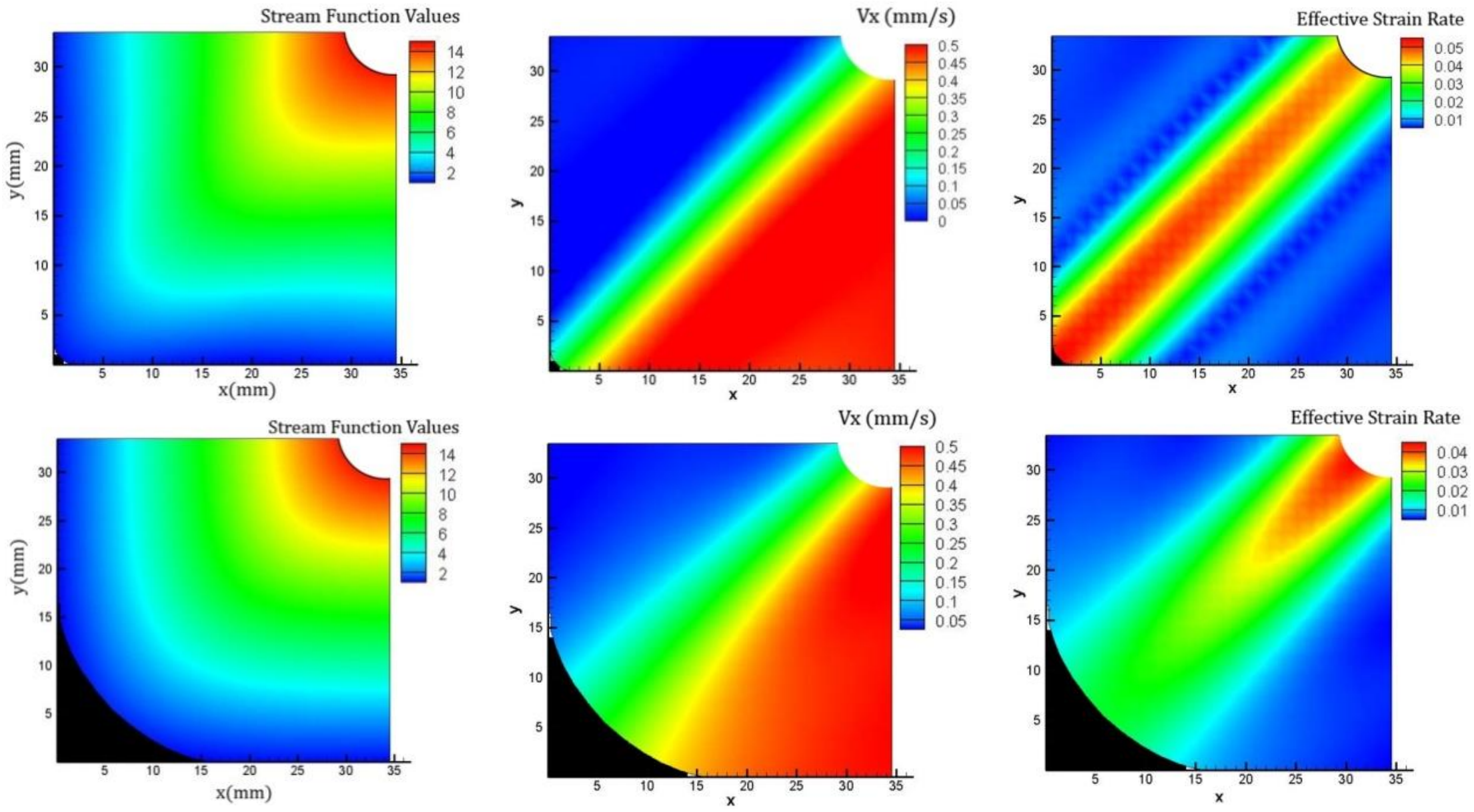

(a) (b) (c)

*Figure 9. Comparison between the predicted contours of a) stream function, b) velocity field and c) effective strain rate, in case of very small and optimized dead metal region during ECAE at 0.5mm/s ram speed.*

It is shown that in contrast with the optimized condition where the velocity field changes gradually, the velocity field alters sharply within the small DMZ case. Accordingly, the calculated effective strain rate contours show severe strain rate magnitudes near the sharp DMZ, while minor strain rates are observable adjacent to the DMZ for optimized state. In other words, the optimization algorithm has tried to lessen the strain rate magnitude at the vicinity of dead metal region that is in accordance with the real world observations. According to the fact that the strain rate within the dead metal region is nearly zero, the model response seems to be rational.

Figure 10 illustrates the flow stress contours calculated at each individual element supposing isothermal situations.

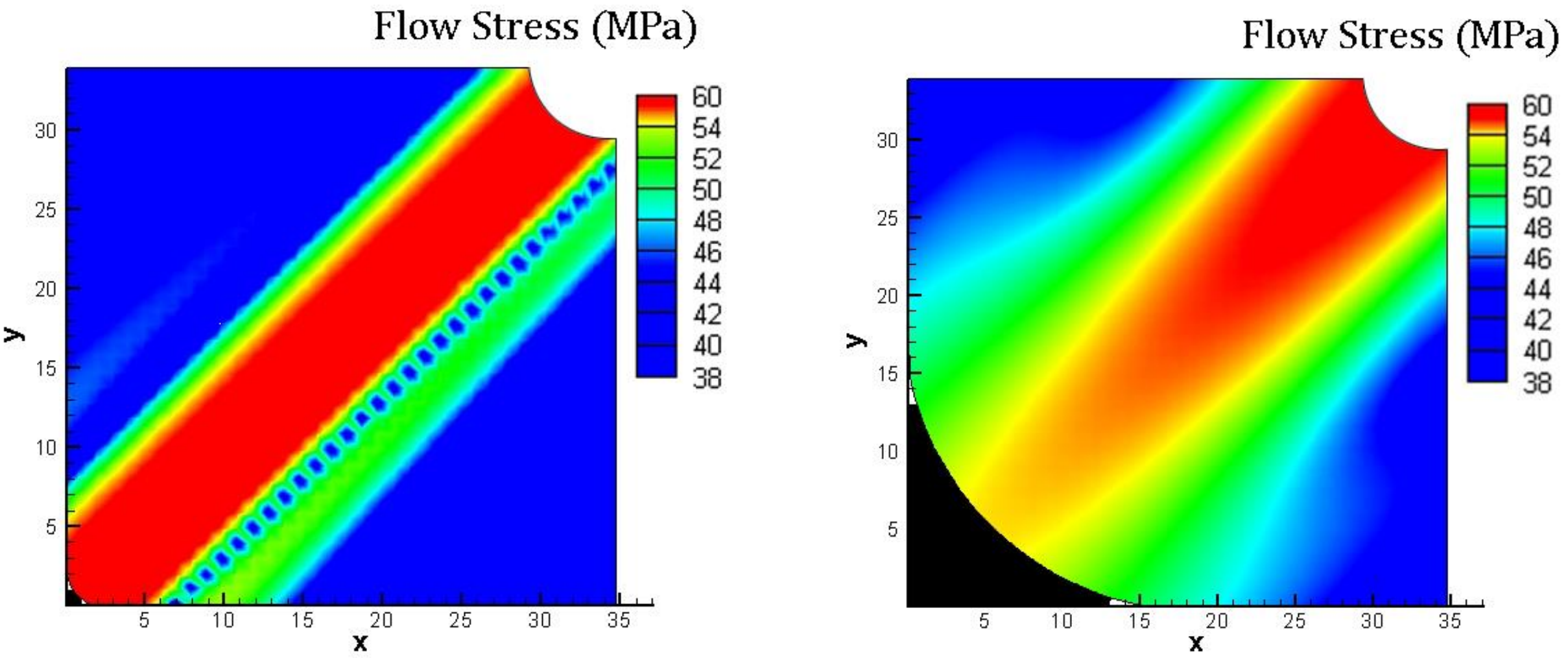

(a) (b)

*Figure 10. Comparison between the calculated flow stress within the deformation zone during ECAE at T=400°C and v=0.5mm/s for a) very small dead metal region, b) optimized dead metal region.*

As can be seen for small DMZ, the maximum flow stress is concentrated on a narrow region while this region is wider within the deformation region for optimized configuration. The above manner will let the internal deformation to roll out, thus, an optimum distribution for the power consumption is accessible.

Figure 11 illustrates the distribution of internal power dissipation per unit area. This quantity is only influenced by the multiple effect of the strain rate and flow stress.

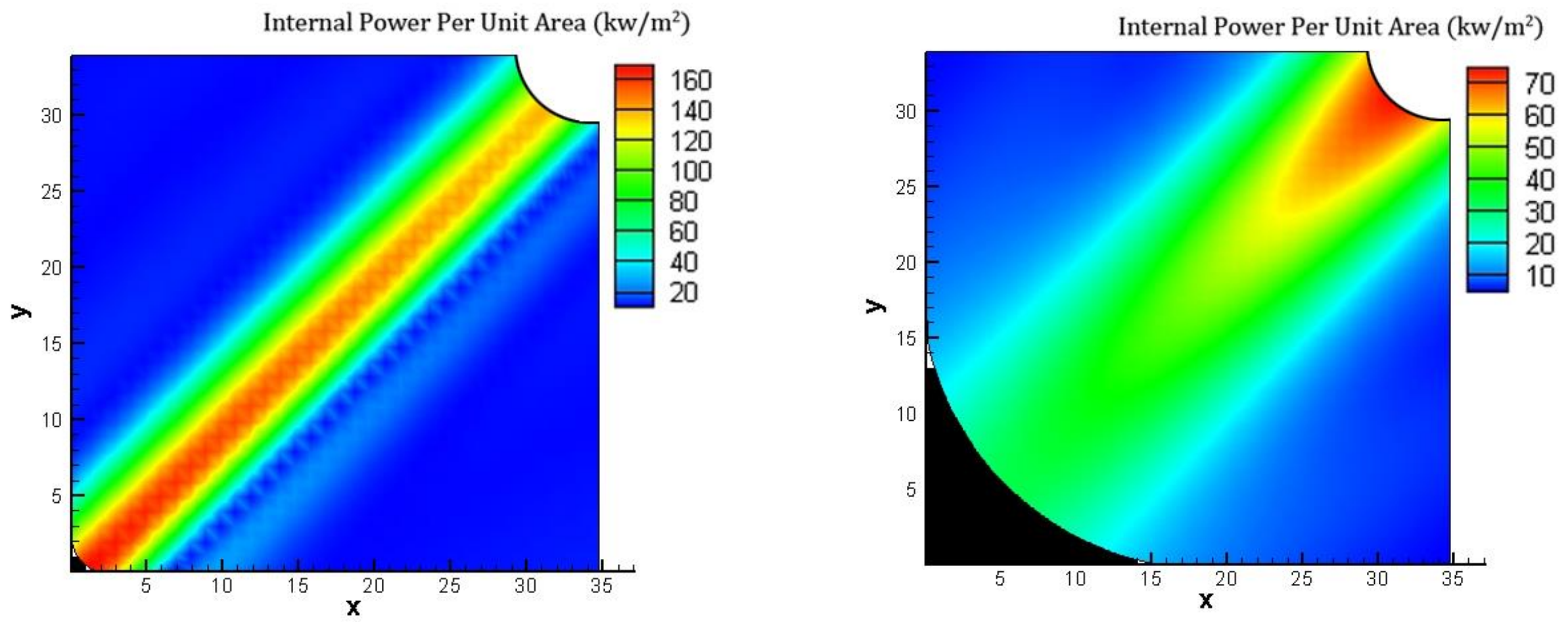

(a) (b)

*Figure 11. Contours of the internal power dissipation per unit area during ECAE at 0.5mm/s ram speed and 400°C for a) very small dead metal region, b) optimized dead metal region.*

It can be inferred that a better distribution of the internal power consumption is attainable within the optimized configuration so that the maximum power consumption occurs far from the dead metal region. In addition, the range of the internal power consumption at the small DMZ situation is veritably greater than the range at the optimized configuration as was expected.

In contrast with many analytical velocity fields that consider unrealistic velocity discontinuity surfaces at the entry and exit of the deformation region [21, 22], at the present work, the deformation field boundaries will be considered alongside the steady state region where no velocity discontinuity is expected. As discussed earlier, the volume constancy rule is remained, which is because of specific definition of the stream function. In return, the power consumption would be infinitesimal close to the steady state region and gradual variations on the velocity field and thus internal power consumption would be discovered by the neural networks by an automatic approach.

Finally yet importantly, the most promising achievement of this approach lies on its capability to fast estimation of the force requirements and deformation field for complex forming processes where no analytical velocity field is provided, neither finite element method works well. Since this subject has not been well investigated so far, it can be a starting point for future investigations. In spite of the achievements, the proposed model exhibits tolerances in its response case by case. As noted, because of the accidental nature of the neural network scheme, observation of some small tolerances for repeated runs of an individual problem is inevitable. In these cases, decreasing the convergence criterion can lead to more accurate results. However, it may rise the solution time substantially. Therefore, consideration of modified algorithm to improve the quality of predicted streamline contours is helpful for rapid assessment of the trained network and can be the subject of next studies. It seems that the proposed approach can be implemented for a wide range of steady state processes as well where an adequate number of training points are accessible. Pursuing the applicability of the proposed approach to simulate the deformation field among strain hardening materials and for other forming processes is an interesting subject for next researches.

## 5. Conclusions

A new algorithm for streamline field generation was introduced based on artificial neural networks implementation. The proposed approach implemented for prediction of the flow pattern and the deformation field during the ECAE process under hot deformation conditions. In this regard, a neural network with specific architecture trained using selected points located at the steady state deformation regions and at the boundary conditions. The network eventually utilized to interpolate the stream function values at the deformation region. Accordingly, the power dissipation components were computed inside the deformation region and the Hill's minimum principle theorem was used to estimate the actual power requirements. A thermo-mechanical finite element code was separately utilized to verify the accuracy of the predicted force requirements and temperature suppositions during the process. Good agreement was observed between the results. Finally, outstanding aspects of the proposed model were taken into account and discussed in detail. The most promising aspect of the proposed approach is the usefulness of this idea in fast computerized simulations that use the upper-bound technique for estimating the force and power requirements for material forming processes.

## Acknowledgement

The author appreciates Dr. Kohbor and Dr. Fattahi for reviewing the text and providing useful comments on the work.